\documentclass[twocolumn]{aastex63}
\usepackage{amsmath}
\usepackage{amssymb}
\usepackage{natbib}
\usepackage{hyperref}
 \usepackage{graphicx}

\usepackage{ulem}

\submitjournal{ApJL}
\accepted{September 29, 2019}

\shorttitle{MHD Seismology of Solar Active  Regions}
\shortauthors{Anfinogentov, S.A. and Nakariakov, V.M.}

\begin{document}
\title{Magnetohydrodynamic Seismology of Quiet Solar Active Regions}

\author[0000-0002-1107-7420]{Sergey~A. Anfinogentov}
\affil{Institute of Solar-Terrestrial Physics SB RAS, Lermontov St. 126, Irkutsk 664033, Russia}
\author[0000-0001-6423-8286]{Valery~M. Nakariakov}
\affil{Centre for Fusion, Space and Astrophysics, Physics Department, University of Warwick, Coventry CV4 7AL, UK}
\affil{St Petersburg Branch, Special Astrophysical Observatory, Russian Academy of Sciences, St  Petersburg, 196140, Russia}

\begin{abstract}
  The ubiquity of recently discovered low-amplitude decayless kink oscillations of plasma loops allows for the seismological probing of the corona on a regular basis.
  In particular, in contrast to traditionally applied seismology which is based on the large-amplitude decaying kink oscillations excited by flares and eruptions, decayless oscillations can potentially provide the diagnostics necessary for their forecasting. We analysed decayless kink oscillations in several distinct loops belonging to active region NOAA 12107 on 10 July 2010 during its quiet time period, when it was observed on the West limb in EUV by the Atmospheric Imaging Assembly on-board  Solar Dynamics Observatory. The oscillation periods were estimated with the use of the motion magnification technique. The lengths of the oscillating loops were determined within the assumption of its semicircular shape by measuring the position of their foot-points. The density contrast in the loops was estimated from the observed intensity contrast accounting for the unknown spatial scale of the background plasma. The combination of those measurements allows us to determine the distribution of kink and Alfv\'en speeds in  the active region. Thus, we demonstrate the possibility to obtain seismological information about coronal active regions during the quiet periods of time.
\end{abstract}

\section{Introduction}
\label{sec:intro}
Active regions of the solar corona are regions of the enhanced plasma density penetrated by a closed magnetic field. In the Extreme Ultraviolet (EUV) band, active regions are seen  as localised bundles of bright plasma loops that are believed to highlight certain magnetic flux tubes. Active regions are known to host sporadic impulsive energy releases observed as solar flares and coronal mass ejections which are the most powerful physical phenomena in the solar system. Robust forecasting of flares and mass ejections is an important element of space weather research. The key required parameter is the magnetic field. But, the direct observational measurement of the coronal magnetic field is possible in some specific cases only, because of the intrinsic difficulties connected with the high temperature and low concentration of the coronal plasma. One promising indirect method for obtaining information about the coronal magnetic field is magnetohydrodynamic (MHD) seismology, based on the estimation of the coronal Alfv\'en speed \citep[e.g.][]{2001A&A...372L..53N, 2014SoPh..289.3233L, 2016GMS...216..395W}. Similar plasma diagnostic techniques are used in laboratory plasma and Earth's magnetospheric research \citep[e.g.,][respectively]{2002PPCF...44B.159F, 2016SSRv..200...75N}.

A suitable seismological probe of the Alfv\'en speed in an active region is a kink (transverse) oscillation of a coronal loop \citep[e.g.][]{1984ApJ...279..857R}. Kink oscillations are excited by low-coronal eruptions, and decay in several oscillation cycles \citep[e.g.][]{2015A&A...577A...4Z}. The spatially-resolving detection of the kink oscillation allowed for the interpretation of the oscillation as the fundamental harmonic of a standing $m=1$ fast magnetoacoustic mode of the coronal loop. The plausibility of this estimation is confirmed by the observationally established linear scaling of the kink oscillation period with the loop length \citep{2016A&A...585A.137G}, and the variety of the oscillation periods detected in different loops belonging to the same bundle \citep{2017ApJ...842...99L}. The first seismological estimation of  the magnetic field in a coronal loop by a kink oscillation was performed by \cite{2001A&A...372L..53N}. The ratio of the wavelength that for the fundamental harmonic is double the length of the loop, and the observationally determined oscillation period gives the phase speed. As the wavelength is much longer than the minor radius of the oscillating loop, it is possible to use the theoretical estimation of the phase speed as the kink speed \citep{1976JETP...43..491R, 1983SoPh...88..179E}. Together with the independent estimation of the density contrast in the loop, this quantity gives the estimation of the local Alfv\'en speed. If there is an independent estimation of the plasma density in the loop, one gets the estimation of the absolute value of the field. An important advantage of MHD seismology by kink oscillations is a clear association of the observed oscillation with a specific plasma structure, which makes the diagnostics free of the line-of-sight integration shortcomings. Moreover, seismology allows for estimating the Alfv\'en speed and field in off-limb active regions where the field could not be determined by extrapolation. 

In addition, the detection of multi-modal kink oscillations has led to the development of kink-based seismological techniques for the estimation of the relative density stratification, based on the ratios of the periods of different harmonics \citep[e.g.][]{2005ApJ...624L..57A, 2007A&A...473..959V}. The transverse profile of the density in the loop could be estimated with the use of another observable parameter, the damping time \citep[e.g.][]{2004ApJ...606.1223V}. Serious progress in coronal seismology by kink oscillations has recently been achieved with the application of  the Bayesian statistics \citep{2015ApJ...811..104A, 2018ApJ...863..167G}, see, also, \citep{2018AdSpR..61..655A}\ for a recent review. An important tool for testing the theoretical results against observations is forward modelling of observables \citep{2016ApJS..223...23Y, 2016ApJS..223...24Y}. The observed combination of two damping regimes, the exponential and Gaussian regimes \citep{2012A&A...539A..37P}, allowed for the development of seismological techniques for the estimation of the transverse profile of the plasma density in the oscillating loop \citep{2018ApJ...860...31P, 2016A&A...589A.136P,2019FrASS...6...22P}, and its evolution in the course of the oscillation \citep{2018ApJ...863..167G}. Certain theoretical shortcomings of the latter techniques have recently been discussed in \citep{2019A&A...622A..44A}. However, the main disadvantage of the seismology by decaying kink oscillations is their occurrence after an impulsive energy release, usually the low coronal eruption \citep{2015A&A...577A...4Z}, which excites the oscillations. This intrinsic difficulty does not allow for the diagnostics of the plasma before the eruption, which would be of interest in the context of space weather forecasting. 

Another, decayless regime of kink oscillations was discovered by \cite{2012ApJ...751L..27W}. Oscillations of this kind are a ubiquitous and persistent feature of \lq\lq quiet\rq\rq\ active regions \citep{2013A&A...560A.107A, 2015A&A...583A.136A}, i.e. they appear in the non-active periods of time. Typical oscillation periods are from a few to several minutes. The periods are found to scale linearly with the length of the oscillating loop, justifying their interpretation as standing kink modes of coronal loops. Moreover, \cite{2013A&A...552A..57N} demonstrated that the same loop oscillates in different periods of time in both decay and decayless regimes with the same oscillation period.
Oscillations of this type are possibly detected in flaring loops too \citep{2018A&A...617A..86L}, and could explain persistent oscillatory variations of the Doppler shift detected in EUV spectral observations by \citet{2012ApJ...759..144T}.
The mechanisms responsible for the sustainability of the oscillations, i.e., counteracting the damping by, e.g., resonant absorption, and hence determining the oscillation amplitude are still debated \citep[e.g.][]{2014ApJ...784..103H, 2015A&A...576A..22M, 2016A&A...591L...5N, 2016ApJ...830L..22A, 2019ApJ...870...55G, 2019FrASS...6...38K}. An intrinsic difficulty in the observational study of decay-less kink oscillations is that their typical projected displacement amplitudes are lower than 1\,Mm, and often smaller than the pixel size of available EUV imagers. Nevertheless, these oscillations are robustly detected with the use of the recently designed motion magnification technique \citep{2016SoPh..291.3251A}. In particular, with the use of this technique, the coexistence of the fundamental and second spatial harmonics of decay-less kink oscillations has been revealed in \citep{2018ApJ...854L...5D}. 
The persistent occurrence of decay-less kink oscillations in coronal active regions before flares and eruptions makes them a promising seismological tool that can provide us with important input parameters for space weather forecasting techniques. 

In this paper, we present the first seismological diagnostics of the Alfv\'en speed in an active region during a non-flaring period of time, i.e., in a quiet active region. 

\section{Observations}
\label{sec:data}
\begin{figure}
    \begin{center}
    \includegraphics[width=0.95\linewidth]{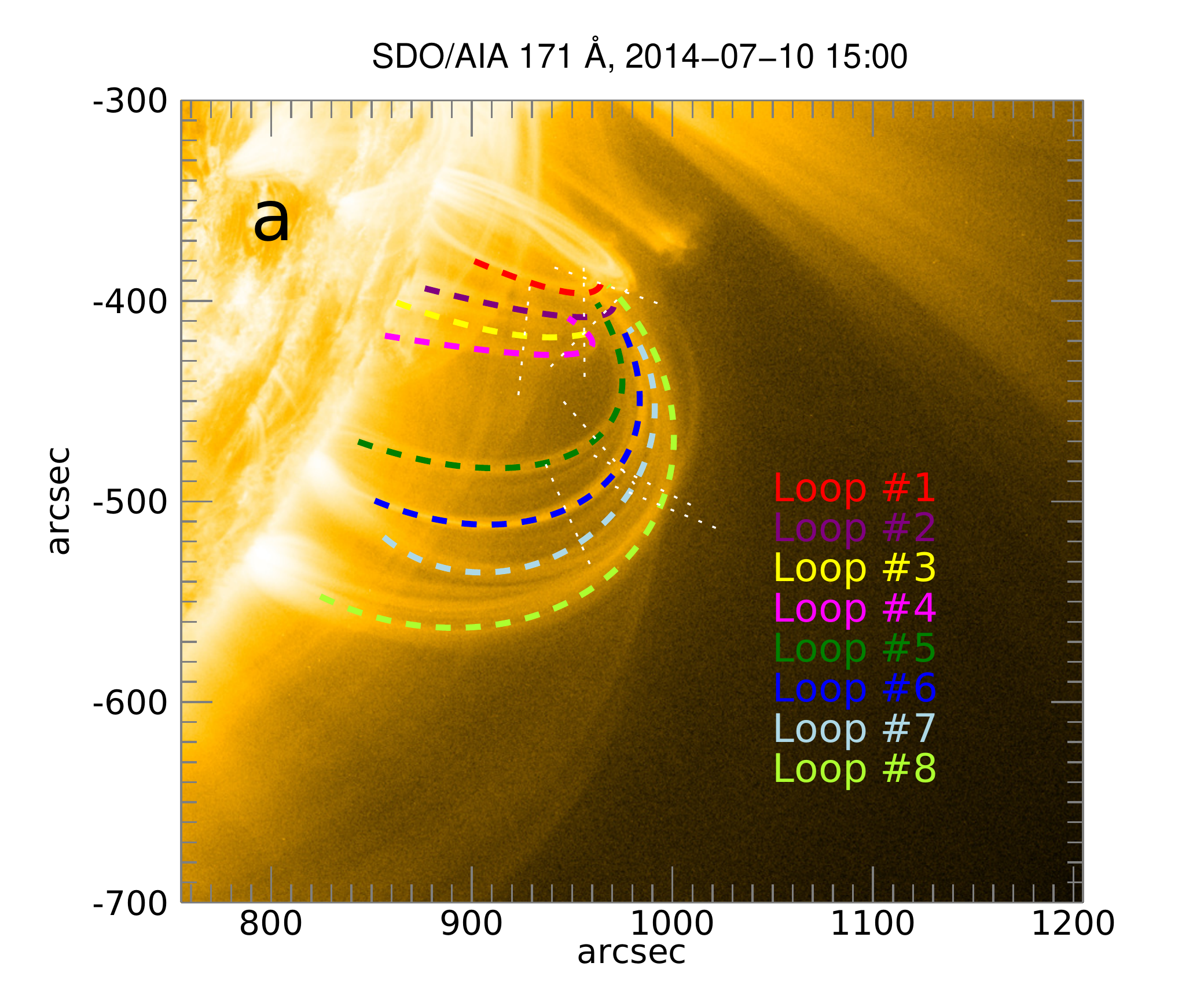}
    \end{center}
    \caption{
    The EUV image of AR 12107 observed by SDO/AIA at 171\,\AA\  on 10 July 2014 at 14:32~UT (background).
    Coronal loops selected for the analysis are over-plotted with the coloured dashed lines.
    Artificial slits used for creating time-distance plots are marked with the straight dotted lines.}
    \label{fig:loops}
\end{figure}

For our study, we selected active region NOAA 12107  observed  on the West limb of the Sun on 10 July 2010.
The active region was seen as a set of coronal loops of different heights, lengths and orientations. Several loops are seen to be  well contrasted in the 171\,\AA\ channel.
We use a 3 hours series of SDO/AIA images recorded from 14:00 UT till 17:00 UT.
No flares or eruptions were observed in the active region or its vicinity during this time interval.
The images were downloaded from the SDO data processing centre (\url{http://jsoc.stanford.edu/}) with the use of the provided on-line service for cutting out the region of interest.
As the analysed active region was located on the solar limb, there was no need for its tracking or derotation.
For the detailed analysis, we selected eight coronal loops indicated in Figure~\ref{fig:loops} with dashed lines of different colours.
In Figure~\ref{fig:loops}, we show an EUV image of the active region NOAA 12107  taken on 10 July 2014 at 14:32 UT. 

\section{Data analysis}

\subsection{Detecting oscillations using motion magnification}

Decayless kink oscillations are obseved to have the displacement amplitude of the order  of 0.2\,Mm \citep{2015A&A...583A.136A} which is less than the pixel size of SDO/AIA.
The analysis of the osillations was performed by processing the imaging data cubes using the motion magnification technique \citep{2016SoPh..291.3251A} based on the Dual Complex Wavelet Transform (D$\mathbb{C}$WT).
Each image is decomposed into a set of complex wavelet components corresponding to different spatial scales, positions and orientations.
The phase of the complex wavelet coefficients reflects the spatial location of different structures in the image and is sensitive to very small displacements of these structures in the next image.
So, the algorithm tracks variations of the phase, and amplifies it in a certain broad range of periods.
Performing the inverse D$\mathbb{C}$WT, we obtain a new series of images where all spatial displacements are magnified by a prescribed factor which is called the magnification coefficient.

In this work, we use the magnification coefficient of 5.
 We found this value optimal for our data-set, since it makes the transverse oscillations well visible in time-distance maps in the well-contrasted loops  on one hand, and does not introduce significant distortion to the images on the other hand.

\subsection{Time-distance maps}

\begin{figure*}
\begin{center}
 \includegraphics[width=1.\linewidth]{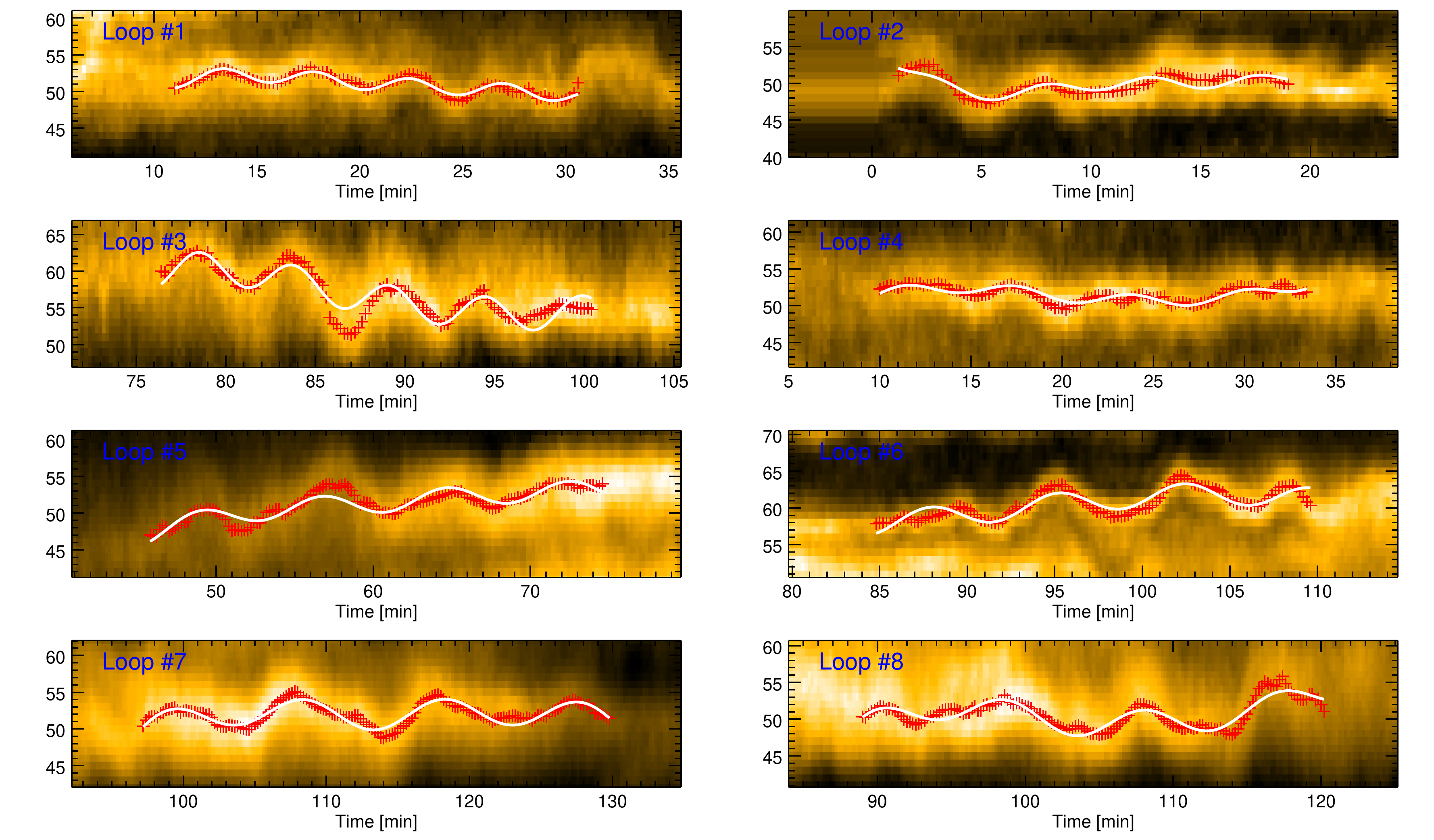}   
\end{center}
\caption{
The time distance plots show decay-less kink oscillations observed in eight loops selected in AR~12107. The kink (transverse) motions in the image sequence were magnified by the factor of 5. The instant of time when the loops appear to be best contrasted in 171\,\AA\ images are marked with vertical green lines. Images taken at these times were used as a reference to estimate the length of the loops.}
\label{fig:td_plots}
\end{figure*}

To make a time-distance map for an oscillating loop, we choose the instance of time where the  loop has the best contrast in the 171\,\AA\ channel, and put an artificial slit across the loop near its apex.
The  slit position and width were manually selected individually for each loop to make the observed oscillation more evident.
The slit positions are indicated in Figure~\ref{fig:loops}, and their widths are listed in Table~\ref{tbl:loops}. In Figure \ref{fig:td_plots}, we show time-distance maps obtained with the use of this technique for the selected loops. The motion magnification allowed us to make the oscillatory patterns clearly visible in time distance maps for all eight loops.

\subsection{Estimation of the density contrast using Bayesian inference}
\label{density}
To estimate the density contrast inside and outside the oscillating loop, we assume that the loop and its neighbourhood are isothermal, and the observed emission is optically thin. Thus, we model the density profile by a step function
\begin{equation}
    {\textstyle n( r) =\begin{cases}
n_{0} , & r< l_{0}\\
n_{\mathrm{e}} , & l_{0} < r< l_{\mathrm{e}}\\
0, & r >l_{\mathrm{e}}
\end{cases}},
\end{equation}
where $n_0$ and $n_\mathrm{e}$ are number densities inside and outside the loop, respectively; $l_0$ is the column depth of the loop segment, connected with the minor radius of the loop, and $l_\mathrm{e}$ is the column depth of the background plasma along the line of sight. The value of
$l_\mathrm{e}$ is expected to be much longer than the minor radius of the loop, i.e., comparable to the active region size.
Note that the estimation of the kink speed requires the value of the external number density $n_e$ in the vicinity of the loop.
The parameter $l_\mathrm{e}$ is introduced to account for the emitting plasma located along the line of sight ahead and behind the oscillating loop.

Within this model the EUV intensity of the loop, $I_0$, and the background, $I_\mathrm{e}$, are calculated as
\begin{equation}
    I_\mathrm{e} = G(\lambda,T) l_\mathrm{e} (\eta n_0)^2,
    \label{eq:int_e}
\end{equation}
\begin{equation}
    I_0 = G(\lambda,T) \left[ (l_\mathrm{e} - l_0) (\eta n_0)^2 + l_0 n_0^2\right],
    \label{eq:int_0}
\end{equation}
where $\eta = n_\mathrm{e}/n_0$ is the density contrast and  $G(\lambda,T)$ is the contribution function that depends upon the observed wavelength and the temperature of the emitting plasma, and accounts for specific properties of the instrument (e.g. SDO/AIA).
In our case, we model only the dependence of the intensity contrast upon the density contrast and, therefore, are not interested in the absolute values.
Thus, for an isothermal plasma, we can safely take $G(\lambda,T)=1$ and $l_0 = 1$.

To estimate the density contrast from the observed intensities $\mathfrak{I}_0$ and $\mathfrak{I}_\mathrm{e}$, which are the modelled intensities $I_0$ and ${I}_\mathrm{e}$ contaminated by the noise, we use the Bayesian analysis in combination with Markov Chain Monte-Carlo (MCMC) sampling.
In our model, we assume that the measurement errors are normally distributed and independent in different pixels, obtaining the likelihood function,
\begin{equation}
    P(\mathfrak{I}_0,\mathfrak{I}_\mathrm{e}|\theta) = \frac{\exp{\frac{-\left[\mathfrak{I}_\mathrm{e} - I_\mathrm{e}(\theta)\right]^2}{\sigma_\mathrm{e}}} 
    \exp{\frac{-\left[\mathfrak{I}_0 - I_0(\theta)\right]^2}{\sigma_0}}}{2\pi\sigma_\mathrm{e}\sigma_0} ,
\end{equation}
where $\theta = [n_0,\eta,l_\mathrm{e}]$ is the set of free model parameters, $\sigma_0$, and $\sigma_\mathrm{e}$  are the measurement errors, and $I_\mathrm{e}(\theta)$ and $I_0(\theta)$ are the modelled intensities given by Eqs.~(\ref{eq:int_e}) and (\ref{eq:int_0}).
In this work, we use uniform priors with the following ranges: [0,1] for the density contrast $\eta$; [10, 200] for the background length scale $l_\mathrm{e}$; and [0,2] for the normalised internal density $n_0$.

The intensity values $I_\mathrm{e}$ and $I_0$ are obtained from the original SDO/AIA images (before the motion magnification) taken at the times when the oscillatory patterns shown in Figure~\ref{fig:td_plots} have been detected.
The measurement uncertainties $\sigma_0$, and $\sigma_\mathrm{e}$ are estimated using the \texttt{AIA\_BP\_ESTIMATE\_ERROR} function from the SolarSoft package \citep{1998SoPh..182..497F}.
Both the intensities and the corresponding errors were then  normalised to such as $I_0=1$. 

To sample the posterior probability distribution, we use our Solar Bayesian Analysis Toolkit (SoBAT) code which is available online at \url{https://github.com/Sergey-Anfinogentov/SoBAT}.
The description of the code can be found in \cite{2017A&A...600A..78P}.
For each analysed loop, we generated $10^6$ samples and used them to find the most probable value of $\eta$. 
The credible intervals are defined as 5 and 95\,\% percentiles and correspond to the confidence level of 90\,\%.

\subsection{Estimating the Alfv\'en speed }
\label{alfven}
Firstly, we estimate the position of the oscillating loop at each instant of time by fitting a Gaussian to the transverse intensity profile of the loop extracted from the time-distance map.
To estimate the period and the corresponding uncertainties from the obtained data points, we use the Bayesian analysis.
Transverse displacements of each loop were modelled by a sinusoidal function on top of a polynomial trend. The measurement errors are assumed to be normally distributed and individually independent.

We generate $10^6$ samples from the posterior distribution using the SoBAT MCMC code.
For all free parameters we use uniform priors.
The kink speed $C_\mathrm{k}$ is then estimated from the oscillation period,
\begin{equation}
C_\mathrm{k} = \frac{2L}{P},
\end{equation}
where  $L$ is the length of the oscillating loop estimated by the apparent position of the loop footpoints and its apex in the assumption of the semicircular shape of the loop. To account for the uncertainties coming from the period measurements we computed $C_\mathrm{k} $ for each of  $10^6$ samples from the posterior distribution of the oscillation period $P$.
It allows us to estimate the most probable value and the credible intervals for  the kink speed, and transparently trace the propagation of the estimated uncertainties in the estimation of the Alfv\'en speed.

The kink speed and density contrast allow us to estimate the external and internal Alfv\'en speeds as 
\begin{eqnarray}
 C_{\mathrm{A0}} &=& C_\mathrm{k}/\sqrt{2/(1+\eta)},\\
C_\mathrm{Ae} &=& C_\mathrm{A0}/\sqrt{\eta},
\end{eqnarray}
respectively \citep[e.g.][]{2001A&A...372L..53N}. The credibility of this technique was demonstrated by \cite{2013ApJ...767...16V} for decaying kink oscillations. 
To account for the uncertainties coming from the measurements of the density contrast $\eta$ and the oscillation period, we calculate $C_{\mathrm{Ae}}$ and $C_{\mathrm{A0}}$  for each of $10^6$ samples obtained with MCMC  for the density contrast $\eta$ and the oscillation period $P$. The most probable values of $C_{\mathrm{Ae}}$ and $C_{\mathrm{A0}}$ are defined as the maximums of the corresponding histograms, and the 90\% credible intervals are calculated as  5\% and 95\% percentiles.

\begin{deluxetable*}{ccccccccc}
\tablecaption{Estimation of the Alfv\'en speed by decay-less kink oscillations}
\label{tbl:loops}
\tablehead{
        \colhead{Loop} & 
        \colhead{Loop}&
        \colhead{Slit }&
        \colhead{Period} &
        \colhead{Intensity} &
        \colhead{Density} &
        \colhead{Kink}&
        \colhead{$C_{A0}$}&
        \colhead{$C_{Ae}$}\\ 
        \colhead{No}
        & \colhead{length [Mm]}
        & \colhead{width [px]}
        & \colhead{[s]}
        & \colhead{contrast}
        & \colhead{contrast}
        & \colhead{speed [km/s]}
        & \colhead{[km/s]}
        & \colhead{[km/s]}
}
\startdata
        1 & 224 & 1 & $276^{+2.8}_{-2.5}$& 0.23& $0.04^{+0.35}_{-0.03}$& $1622^{+15}_{-17}$& $1173^{+182}_{-23}$& $4313^{+7935}_{-2156}$ \\
        2 & 231 & 5& $334^{+40}_{-49}$& 0.46& $0.07^{+0.40}_{-0.05}$& $1395^{+226}_{-163}$& $942^{+338}_{-35}$& $2765^{+4221}_{-1076}$ \\
        3 & 244 & 11&$321^{+11}_{-7.8}$& 0.66& $0.11^{+0.52}_{-0.04}$& $1525^{+38}_{-52}$& $1140^{+240}_{-38}$& $2122^{+2156}_{-384}$ \\
        4 & 235 & 5& $382^{+18}_{-15}$& 0.70& $0.12^{+0.57}_{-0.04}$& $1228^{+49}_{-58}$& $927^{+218}_{-43}$& $1549^{+1514}_{-184}$ \\
        5 & 292 & 28&  $475^{+10}_{-10}$& 0.50& $0.08^{+0.42}_{-0.06}$& $1229^{+26}_{-25}$& $903^{+161}_{-27}$& $1974^{+3578}_{-466}$ \\
        6 & 329 & 5& $435^{+12}_{-11}$& 0.43& $0.07^{+0.43}_{-0.05}$& $1512^{+39}_{-42}$& $1110^{+201}_{-38}$& $2624^{+5352}_{-769}$ \\
        7 & 343 & 15&$580^{+6.7}_{-6.6}$& 0.42& $0.06^{+0.43}_{-0.04}$& $1184^{+14}_{-14}$& $866^{+155}_{-21}$& $1948^{+4051}_{-489}$ \\
        8 & 391 & 13& $547^{+9.0}_{-8.5}$& 0.26& $0.04^{+0.37}_{-0.03}$& $1429^{+22}_{-23}$& $1030^{+174}_{-19}$& $3353^{+6682}_{-1478}$  \\
\enddata
\end{deluxetable*}

\subsection{Mapping the Alfv\'en speed in the corona}
The detection of kink oscillations in different coronal loops with different heights and lengths allows us to make spatially resolved estimates of the Alfv\'en speed in the active region.
The estimation requires the knowledge of the loop lengths, oscillation periods corresponding to the fundamental kink mode, and the density contrasts in the oscillating loops \citep[see][and Section~\ref{alfven}]{2001A&A...372L..53N}.
The oscillation period is estimated directly from the time-distance maps, while the observed intensity contrast inside and outside the oscillating loop gives us a proxy for the density contrast.
The length of the oscillating loop is estimated from the  position of its foot-points.

\begin{figure}

\begin{center}
\includegraphics[width=0.9\linewidth]{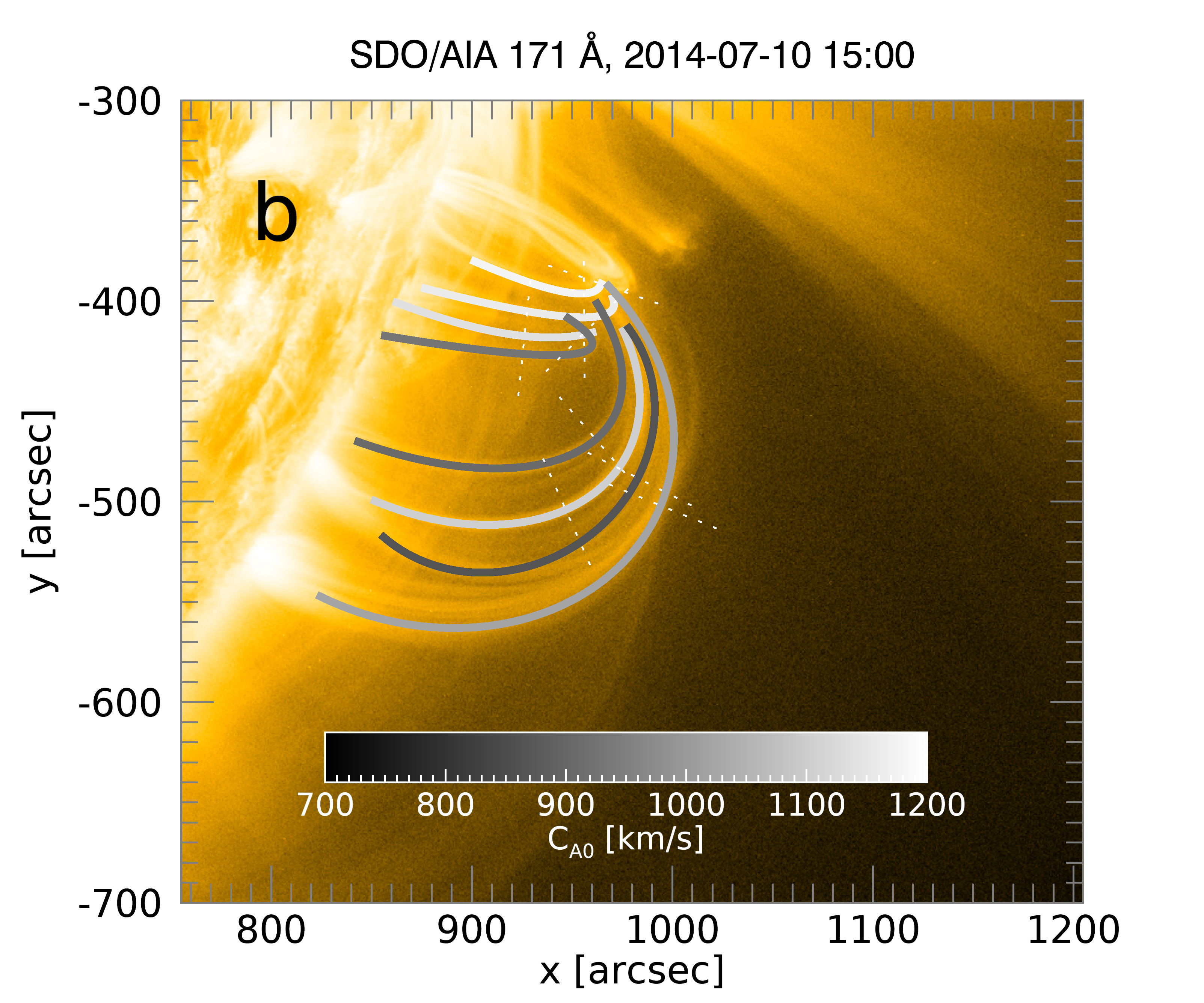}
\end{center}
\caption{Mapping the Alfv\'en speed in AR 12107.
 The colour of the broad curved lines following the coronal loops show the internal Alfv\'en speed $C_\mathrm{A0}$  estimated from the observed decay-less kink oscillations.
The EUV image of AR 12107 observed by SDO/AIA at 171\,\AA\  on 10 July 2014 at 14:32 UT is used as the background.
 Artificial slits used for creating time-distance plots are marked with the straight dotted lines.}
\label{fig:Ca_map}
\end{figure}

For the eight chosen coronal loops, we estimated the internal and external Alfv\'en speeds and the corresponding uncertainties.
Our estimates are summarised in Table~\ref{tbl:loops}. Note, that despite the huge uncertainty in the density contrast estimations, we successfully obtained reliable measurements of the Afv\'en speed inside oscillating loops with the accuracy of 15-20\%.
In Figure \ref{fig:Ca_map}, we show a spatially resolved mapping of the internal Alfv\'en speed inferred from the decayless kink oscillations. The given values  should be understood as values averaged along the oscillating loops, therefore the colour corresponding to the Alfv\'en speed value is evenly distributed along each loop.

\section{Discussion}
\label{sec:conlusion}

We demonstrated that decay-less kink oscillations processed by the pioneering motion magnification technique allow one to map the Alfv\'en speed in the solar corona during the quiet time period. The analysis of the EUV emission from active region 12107 produced the first ever seismogram of  a solar coronal active region during its quiet period, showing the spatial distribution of the  Alfv\'en speed.
The seismogram is presented in Figure~\ref{fig:Ca_map}.
Note, that  the quantities shown in Figure~\ref{fig:Ca_map} correspond to the values averaged along the oscillating loops, as the effect of the plasma stratification in this study was neglected.
However, further accounting for higher spatial harmonics of decayless kink oscillations \citep{2018ApJ...854L...5D} would allow for mapping the Alfv\'en speed along the loops.

Relatively large errors in the estimations of the density contrast from the observed intensity contrast lead to a very uncertain estimate of the Alfv\'en speed in the plasma outside the loops (see Table \ref{tbl:loops}).
However, the Alfv\'en speed inside the oscillating loop can be measured with the precision of about 15--20\,\%.
Even lower uncertainties can be achieved if more precise measurements of the density contrast are available in the same active region either spectroscopically, or by seismology based on decaying kink oscillations \citep[e.g.][]{2016A&A...589A.136P,2018ApJ...860...31P} .
Note that the less uncertain Alfv\'en speed inside the oscillating loop is more informative, since it can be recalculated to the magnetic field strength after measuring independently the plasma density in the loop, while measuring the density of the background plasma is far more complicated.
The density of a coronal loop can be obtained, for example, using the forward modelling approach \citep{2018ApJ...863..167G}, or from the analysis of the differential emission measure \citep[see e.g.][]{2013SoPh..283....5A}.
Even if the robust estimation of the plasma density is not possible, the estimation of the Alfv\'en speed is important for, for example, understanding the interaction of global coronal waves with the active region hosting the oscillating loops \cite[e.g.][]{2017SoPh..292....7L}. In addition, further improvement of the method can be achieved by making more precise the estimation of the length of the oscillating loop \citep[see, e.g.][]{Aschwanden2011}. 

In addition, we demonstrated that decayless kink oscillations can be detected in many loops within a single active region with the use of presently available EUV images provided by SDO/AIA. This detection allows us to carry out spatially resolved measurements of the Alfv\'en speed and, hence, potentially, the coronal magnetic field during the quiet time periods.

We should emphasise, that the coronal seismology based on decayless kink oscillations such as presented here can be performed routinely for almost every active region observed on the Sun, since decay-less kink oscillations are a ubiquitous phenomenon \citep{2015A&A...583A.136A}, and are detected almost always in the most of active regions.

The obtained results may be considered as the first step towards the routine estimation of the Alfv\'en speed and, potentially, of the magnetic field and free magnetic energy available for the release in, in particular, pre-flaring active regions.


\acknowledgments
This work was supported by the Russian Scientific Foundation grant No. 18-72-00144 (S.A.A., Data analysis sections, processing data and preparing figures,interpretation of the obtained results).
V.M.N. (Introduction and Discussion sections, interpretation of the obtained results) acknowledges support from the STFC consolidated Grant No. ST/P000320/1 and from the Russian Foundation for Basic Research grant No. 18-29-21016.
The authors thank the SDO/AIA team.

\facility{SDO/AIA (Atmospheric Imaging Assembly onboard Solar Dynamics Observatory)}

\software{IDL, SolarSoft, DT$\mathbb{C}$WT based motion magnification \citep{2016SoPh..291.3251A,sergey_anfinogentov_2019_3368774}, Python implementation of DT$\mathbb{C}$WT \citep{rich_wareham_2017_889246}, Solar Bayesian Analysis Toolkit}

\section{Competing  interests}
The authors declare that they have no conflicts of interests.

\bibliography{Decayless_seis_intro}
\bibliographystyle{aasjournal}

\end{document}